\documentclass[british, 11pt, a4paper]{article}
\usepackage[left=1in,top=1in,right=1in,bottom=1in,nohead,paperwidth=8.5in, paperheight=11in]{geometry} %1 inch margins, 8 1/2 x 11-inch pages: standard choices for most journals
\usepackage{hyperref}
\usepackage[utf8]{inputenc}
\usepackage{natbib} %standard package for references
\setcitestyle{authoryear,open={(},close={)}}
\usepackage{ntheorem, graphicx, amsmath}
\theoremseparator{:}
\newtheorem{hyp}{Hypothesis}

\begin{document}

\title{Detecting Anti-dumping Circumvention:\\a Network Approach}
\author{
Luca Barbaglia$^{a}$\footnote{Corresponding author: Luca Barbaglia. The views expressed are purely those of the authors and do not, in any circumstance, be regarded as stating an official position of the European Commission. 
{\tt E-mail}: 
luca.barbaglia@ec.europa.eu, 
christophe.croux@edhec.edu, 
i.wilms@maastrichtuniversity.nl.
}, 
Christophe Croux$^{b}$, 
Ines Wilms$^{c}$
%Elisa Tosetti$^{d}$,
\\
%\mbox{}\\
{\small $^{a}$ European Commission, Joint Research Centre}\\
{\small $^{b}$ EDHEC Business School, Lille, France}\\
{\small $^{c}$ Maastricht University, The Netherlands}\\
%{\small $^{d}$ Ca' Foscari University of Venice, Italy}\\
\\ %\mbox{} \\
%{PRELIMINARY: DO NOT CITE}
}

\date{\today}
%\date{}

\date{}
\maketitle
\setcounter{page}{1}

% \vspace{-0.5cm}
% \begin{center}
% \Large{\textsc{Preliminary and incomplete: \\ please do not circulate}}
% \end{center}

\vspace{0.2cm}
\begin{abstract}
\begin{center}
\mbox{}\\
\begin{minipage}{14cm}
\noindent
{\small 
Despite the increasing integration of the global economic system, anti-dumping measures are a common tool used by governments to protect their national economy.
In this paper, we propose a methodology to detect cases of anti-dumping circumvention through re-routing trade via a third country.
Based on the observed full network of trade flows, we propose a measure to proxy the evasion of an anti-dumping duty for a subset of trade flows directed to the European Union, and look for possible cases of circumvention of an active anti-dumping duty.
Using panel regression, we are able correctly classify 86\% of the trade flows, on which an investigation of  anti-dumping circumvention has been opened by the European authorities.
} 
%\mbox{}\\
%\mbox{}
\\
\end{minipage}
\end{center}

\end{abstract}

\newpage

\section*{Non-technical summary}

%% CIRCUMVENTION
The widespread adoption of anti-dumping measures has induced economists to study their effects on trade flows in terms of volume and prices. 
Anti-dumping circumvention generally refers to an attempt by countries subject to anti-dumping to avoid paying the duties by ``formally" moving outside the range of the anti-dumping duties order while ``substantially” engaging in the same commercial activities as before. 
One type of anti-dumping circumvention occurs when exporting products subject to anti-dumping duties from third countries (re-routing).

%% THIS PAPER
In this paper, we adopt a network perspective to build a \textit{circumvention index} that allows us to study possible events of anti-dumping circumvention when focusing on a subset of trade flows directed to the EU. 
We compare the structure of the network in international trade flows before and after the introduction of a specific anti-dumping duty and study the instance of a possible anti-dumping circumvention via re-routing during the period of activity of the duty.
This paper contributes to the existing literature by adopting a systematic approach that does not require the user to manually identify the third country by which circumvention might happen. Relying on the complete network of observed trade flows, we analyse the possibility of an illicit practice happening via any node (i.e., a country) in our network, thus not requiring any prior knowledge or assumption.

%% RESULTS
Our results show that studying the proposed circumvention indexes via a panel regression model allows us to correctly detect 86\% of the trade flows  that have actually been put under investigation by the European authorities with an anti-dumping circumvention allegation.
The proposed methodology could be used as an early-warning signal which might proceed with a more in-depth and thorough analysis of the trade flows patterns using administrative and accounting data.

\newpage

\section{Introduction}

Despite the growth in international trade due to the increased integration of national economies into a global economic system as well as advancements in telecommunications and logistics, we observe persistent and even intensifying adoption of trade protection measures \citep{cerasa2019hedonic, tabakis2019preferential}. 
Anti-dumping intervention is often adopted by governments as a tool to protect domestic firms and industries, with strong implications for their earning management strategies \citep{godsell2017earnings} and for the overall trade volumes \citep{egger2011}.
``Dumping" occurs when a firm charges a lower price for the same product on the foreign market than it does in the domestic market. When it is proved that the dumped imports are damaging the competing industry in the importing country, the latter can impose anti-dumping measures to provide relief to domestic industries injured by imports. The country’s imposition of an anti-dumping duty is determined by the dumping margin, namely, the difference between the export price and the domestic selling price in the exporting country. By adding the dumping margin to the export price, the dumped price can be rendered a fair trade price.

%% FOUCS ON THE EU ANTIDUMPING MEASURES
%https://ec.europa.eu/trade/policy/accessing-markets/trade-defence/actions-against-imports-into-the-eu/anti-dumping/#:~:text=A%20non%2DEU%20company%20is,cost%20of%20production%20and%20profit.
The European Union (EU) has implemented a number of measures to protect its member state's national economies.
The implementation of such trade defence measures aims at protecting European producers from illicit export strategies from commercial partners, as well as final consumers in the EU, as documented by \cite{juramy2018anti}. 
An anti-dumping investigation can be opened by the European authorities by their own initiative or after an official complaint introduced by an European  producer. 
The opening of an anti-dumping investigation (also referred to as proceeding) is published on the EU's Official Journal\footnote{The archive of the EU's Official Journal can be consulted at \url{https://eur-lex.europa.eu/oj/direct-access.html}.}, where, among other issues, it is assessed  (i) whether there is evidence of dumping, (ii) what the damage caused to European producers is, and (iii) whether there exists a causal link between the dumping and the associated economic damage.
%% ADD SOME NUMBER ABOUT EU ANTIDUMPING
We refer to \cite{nita2013first} for a detailed presentation of all steps that lead to the adoption of an anti-dumping duty by the European authorities.

%% CIRCUMVENTION
The widespread adoption of anti-dumping measures has induced economists to study their effects on trade flows in terms of volume and prices, as well as company behaviour (for a recent study on firm-level effects of European anti-dumping duties, see \citealp{felbermayr2020trade}). 
Anti-dumping circumvention generally refers to an attempt by countries subject to anti-dumping to avoid paying the duties by ``formally" moving outside the range of the anti-dumping duties order while ``substantially” engaging in the same commercial activities as before. 
One type of anti-dumping circumvention occurs when exporting products subject to anti-dumping duties from third countries (re-routing). 
%Another type of anti-dumping circumvention occurs through product misclassification, namely reclassify the export to products that have only minor differences with those subject to anti-dumping duties.
\cite{vermulst2015eu} provides an overview of the measures adopted by the EU in order to avoid the circumvention of an anti-dumping duty, advocating for more transparent and predictable anti-circumvention rules adopted at the international level.

%% THIS PAPER
In this paper, we adopt a network perspective to build a \textit{circumvention index} to study possible events of anti-dumping circumvention. We focus on a subset of trade flows directed to the EU. 
While most of the official investigation procedures rely heavily on administrative and accounting data \citep{godsell2017earnings}, researchers have been focusing on trade flow data to detect possible cases of anti-dumping circumvention (e.g., see \citealp{liu2019anti} or \citealp{rousseeuw2019robust}).
We look at the structure of the network in international trade flows before and after the introduction of a specific anti-dumping duty and interpret changes in certain network features as an indication of possible anti-dumping circumvention (for an overview of the application of network theory to model international trade patterns, we refer to \citealp{zhou2016} and \citealp{du2017}).
For our analysis, we use monthly data covering the period from 2000 to 2015 from the United Nations Commodity Trade (Comtrade) statistics database \citep{bown2016a}. 
We match these data with information on anti-dumping measures and investigations extracted from the Global Anti-dumping Database, and study the instance of a possible anti-dumping circumvention via re-routing during the period of activity of the duty.
We provide an out-of-sample assessment of the performance of the proposed methodology against the official investigations opened by the European authorities, which we consider as true labels for a suspicious circumvention event.

%% CONTRIBUTION
Our work relates closely to the anti-dumping circumvention of \cite{liu2019anti}, who also rely on Comtrade data to study events of  circumvention of anti-dumping duties by Chinese exporters.
This paper contributes to the existing literature by adopting a systematic approach that does not require the user to manually identify the third country by which circumvention might happen. Relying on the complete network of observed trade flows, we analyse the possibility of an illicit practice happening via any node (i.e., a country) in our network, thus not requiring any prior knowledge or assumption.
%% RESULTS
Our results show that studying the proposed circumvention indexes via a panel regression model allows us to correctly detect 86\% of the trade flows  that have actually been put under investigation by the European authorities with an anti-dumping circumvention allegation.
The proposed methodology yields an early-warning signal which might be followed by a thorough analysis of the trade flows patterns using administrative and accounting data.

%% STRUCTURE
This paper is structured as follows. 
Section \ref{sec_literature} reviews the existing anti-dumping literature.
Section \ref{sec_data} introduces the various data sources used in the analysis. 
The  network approach to compute anti-dumping circumvention measures is presented in  Section \ref{sec_methods}, as well as the econometric method used for detecting anti-dumping circumvention.
Section \ref{sec_results} discusses the results and 
Section \ref{sec_conclusions} provides some conclusions together with venues for future research.

\section{Literature review\label{sec_literature}}

%\section{Literature Review}
The literature on the economic effects of anti-dumping measures is wide.
We re-address the reader to \cite{bown2016empirical} and \cite{bown2020global} for two empirical studies about the trade policies at the international level and their bilateral impact on global value chains.
\cite{egger2011} use data on international trade flows from the International Monetary Fund to investigate the impact of anti-dumping duties on trade volume, over the period 1960-2001. The authors concluded that the volume and welfare effects have been negative, but quite modest.
\cite{baltagi2015} review a set of estimation techniques for gravity models of bilateral trade of goods, with a specific focus on a range of problems that arise when modelling international trade flows such as issues of endogeneity, estimation with missing or zero trade flow data.
%Anti-dumping measures are often taken in protection of national producers. \cite{juramy2018anti} focuses on the impact of EU anti-dumping policy on final European consumers.

%% FRAUD DETECTION
Numerous works have investigated international trade data for fraud detection purposes. 
\cite{barabesi2016modeling} propose the use of the Tweedie distribution to flexibly model trade quantities and assess the statistical performance of anti-fraud methods.
\cite{rousseeuw2019robust} combine the FastLTS algorithm for robust regression with alternating least squares to detect outliers and level shifts, and show its potential application on two real-life examples. 
\cite{muhammad2019vino} analyse the impact of fraudulent behavior on Chinese wine imports  by combing information on traded quantities and prices at the 6-digits product classification and on fraud mentions in the media. 
\cite{gara2019magic} study inconsistencies in the mirror statistics  at the 6-digits product classification between two partner countries to create an indicator of the potential risk of observing an illegal transaction.

%% FOCUS ON CIRCUMVENTION paper of Liu
Our work is related to the work of \cite{liu2019anti}. These authors rely on custom  data to study the circumvention of the United States (US) anti-dumping measures by Chinese exporters through trade re-routing via third countries.
Their results indicate that the adoption of US anti‐dumping measures is associated to a higher positive correlation between US imports from a third country and Chinese export to the same third country.
The authors show that this positive correlation is stronger for trade flows subject to an anti-duping, which is line with the idea of anti-dumping circumvention through re-routing via a third country.

\section{Data\label{sec_data}}
We join three main data sources: (i) bilateral trade flows from the United Nations Comtrade, (ii) information about anti-dumping measures by the World Bank Global Anti-dumping database, (iii) anti-circumvention investigations by the European Union. The first two data sources provide the full observed network of trade flow and associated anti-dumping duties that we use to build some measures of circumvention of the anti-dumping.

%\subsection{Data sources}
%% COMTRADE
The Comtrade database provides data about import and exports at the 6-digit level in the Harmonised System (HS) classification for all members of the United Nations \citep{comtrade2015comtrade}. 
Comtrade provides the aggregates of quantity exchanged (Kg) and trade value (US Dollars, USD) at monthly frequencies starting from January 2010. 
% Mirror stats
For each record, we observe the trade flow type, that is import or export\footnote{Comtrade provides 4 trade flow types: imports, exports, re-imports, re-exports. For simplicity, consider re-imports (re-exports) as imports (exports).}, and the two partner countries. 
In presence of mirroring trade flows, we average between the reported values (for more details on mirror statistics using Comtrade data, see \citealp{hamanaka2012whose} and \citealp{gara2019magic}).
% Europe as one country
We remove all intra-EU trade flows and consider the EU as a unique trade partner.
Overall, our Comtrade data consists of 205 reporting countries, 6058 commodities for an average number of approximately 1.7 million trade flows (i.e., one commodity traded from country $i$ to country $j$) observed each month.
%We refer to \cite{liu2019anti} for another work on anti-dumping circumvention using Comtrade data.

%% GLOBAL ANTI-DUMPING DATABASE
We augment the data set on trade flows with the World Bank Global Anti-dumping data set \citep{bown2016a}, which provides information about the date of the announcement, imposition and withdrawal of an anti-dumping measure against a partner country with respect to a set of goods from 1980 to 2015.
As for the interest of our application, we focus only on anti-dumping measures imposed by the EU.  
In particular, for 43 trade flows we observe either the imposition date or the revocation date (or both) of an anti-dumping measure by the EU in the period 2010-15.

%% EU ANTI-DUMPING INVESTIGATIONS
In order to properly assess the performance of our detection method, we build a data set containing information about all open and ongoing anti-circumvention investigations ran by the EU.
We scrape the complete history of anti-fraud investigations published on the European Commission Directorate-General for Trade website\footnote{The official investigations are published on the European Commission Directorate-General for Trade website, available at \url{https://trade.ec.europa.eu/tdi/completed.cfm}.}. 
In particular,  for any investigation where we find the specific mention  ``anti-circumvention", we extract the 8-digit commodity code, the countries under investigation and the publication date:
there are 193 instances of investigations about anti-dumping circumvention potential events run by the EU over the period 2013-19.
Grouping the investigation at the 6-digit classification level and matching it with the Comtrade data, we obtain a set of 20 trade flows that were subject to an anti-circumvention investigation by the EU over the period 2010-15. These data are not an input of our models as they are only used as a test data set in order to compare the performance of the proposed method.

%% TIME SERIES in CONSIDERATION: only 18 of them
Our data set consists of a very large number of unique 6-digit trade flows reported among any two countries in the world, to which we attach information about anti-dumping duties and eventual anti-dumping circumvention investigations.
%\footnote{The data set is available on the corresponding author's personal website \url{https://lucabarbaglia.github.io/}.}. 
In order to keep the exploratory analysis within bounds, we decide to investigate the anti-dumping circumvention only on a subset of products imported by the EU.
%% first 2 digits
First, we keep only products with the first two digits belonging to ``68, 69, 70, 72, 73, 81, 83, 84, 85", corresponding to the macro-categories of ceramics, glass-fibres, steel, iron and electric motors and generators.
We focus on this subset of series since they relate to highly capital intensive industries, whose production cannot be flexibly increased in the short run: 
should we observe that a country who is not a large exporter of one of these products experiences a sudden increase in export, this might indicate the potential re-routing of the commodity.
%% Impose and REVOKE
Second, we consider only the time series for which the European authorities imposed and revoked an anti-dumping duty over the period in analysis. 
In this way, we can clearly identify a period where the trade flow is subject to an anti-dumping duty: if a circumvention of an anti-dumping duty happened, we should observe it in this time span.

%% RESULTS 18 TS
As a result, we obtain a subset of 18 commodities of trade flows starting from January 2010 to December 2015, for which we observe the imposition and withdrawal of an anti-dumping duty over the period in analysis.
Table \ref{tab_series} reports the trade flows in analysis, the description of the product and whether the trade flow has been subject to an investigation about anti-dumping circumvention by the European authorities as well as the country of origin.
Relying on the data for official investigations about anti-dumping circumvention, we observe that 14 of the commodities in analysis have been subject of an investigation by the European authorities.

\begin{table}[]
\centering
\caption{Trade flows by HS commodity code, product description, country of origin and investigation: if \textit{yes}, the trade flow has been subjected to an investigation by the European authorities.}
\label{tab_series}
\bigskip
\footnotesize{
\begin{tabular}{
p{0.09\linewidth} 
p{0.6\linewidth} 
p{0.07\linewidth} 
p{0.12\linewidth} }
HS code & Description & Origin & Investigation \\ 
  \hline
690710 & Ceramic tiles, cubes and similar articles; unglazed, whether or not rectangular, the largest surface area of which is capable of being enclosed in a square the side of which is less than 7cm & China & no \\ 
  691110 & Tableware and kitchenware; of porcelain or china & China & yes \\ 
  701911 & Glass fibres; (including glass wool), chopped strands, of a length of not more than 50mm & China & no \\ 
  701940 & Glass fibres; woven fabrics of rovings & China & no \\ 
  701951 & Glass fibres; woven fabrics (other than of rovings), of a width not exceeding 30cm & China & yes \\ 
  701959 & Glass fibres; woven fabrics (other than of rovings), n.e.s. in item no. 7019.5 & China & yes \\ 
  722611 & Steel, alloy; flat-rolled, width less than 600mm, of silicon-electrical steel, grain-oriented & Japan & no \\ 
  730411 & Iron or steel (excluding cast iron); seamless, line pipe of a kind used for oil or gas pipelines, of stainless steel & China & yes \\ 
  730441 & Steel, stainless; cold-drawn or cold-rolled, tubes and pipes of circular cross-section & China & yes \\ 
  730490 & Iron or steel; tubes, pipes and hollow profiles, seamless, n.e.s. in heading no. 7304 & China & yes \\ 
  810296 & Molybdenum; wire & China & yes \\ 
  850131 & Electric motors and generators; DC, of an output not exceeding 750W & China & yes \\ 
  850132 & Electric motors and generators; DC, of an output exceeding 750W but not exceeding 75kW & China & yes \\ 
  850133 & Electric motors and generators; DC, of an output exceeding 75kW but not exceeding 375kW & China & yes \\ 
  850134 & Electric motors and generators; DC, of an output exceeding 375kW & China & yes \\ 
  850162 & Electric generators; AC generators, (alternators), of an output exceeding 75kVA but not exceeding 375kVA & China & yes \\ 
  850163 & Electric generators; AC generators, (alternators), of an output exceeding 375kVA but not exceeding 750kVA & China & yes \\ 
  850164 & Electric generators; AC generators, (alternators), of an output exceeding 750kVA & China & yes \\ 
   \hline
\end{tabular}
}
\end{table}

\begin{figure}
    \centering
    \includegraphics[scale=0.7]{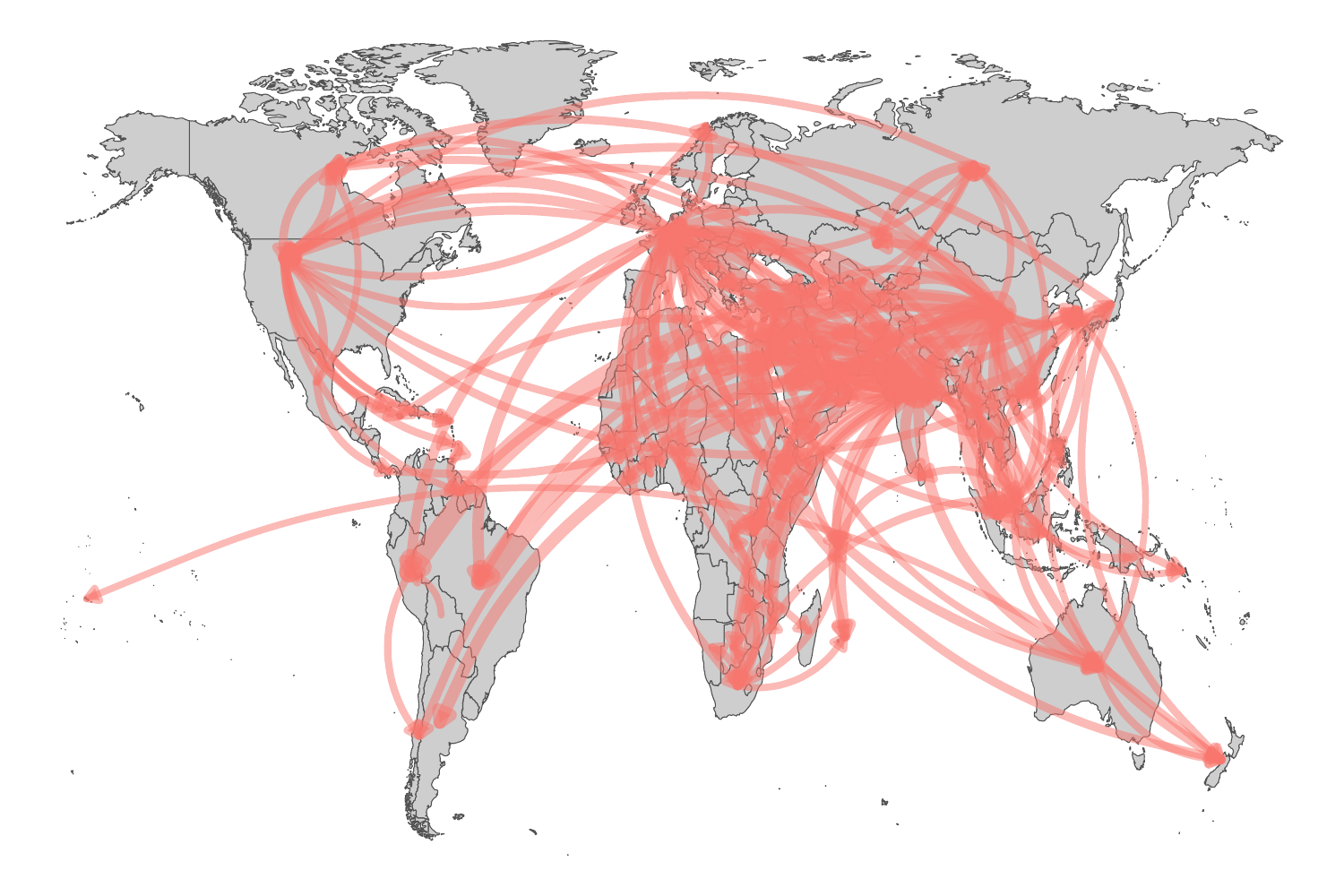}
    \caption{Network of the observed trade flows for commodity code 690710 in December 2015. A directed edge indicates the direction of the trade flow going from the exporting country to the importer. The edge size is proportional to the monetary value exchanged. }
    \label{fig_trade_net}
\end{figure}

%% EXAMPLE OF A TRADE NETWORK
As an illustration, Figure \ref{fig_trade_net} presents the observed network of trade flows for product 690710 in the last month in our analysis, that is December 2015. 
The network is weighted and directed, where the edge direction goes from the exporting to the importing countries, and the edge size is proportional to the monetary value of the trade flow. 
Out of the 205 unique countries represented in our data set, only 94 of them reported a trade flow of product 690710 in December 2015. 
Focusing on the trade flows directed to the EU, we observe that they are originating from 12 distinct countries, which also imported the same product from 32 third countries in the same month.

\section{Methods\label{sec_methods}}

In this section, we  present our empirical strategy to study events of anti-dumping circumvention via re-rerouting a trade flow via a third country.

\subsection{Network-based circumvention measures}
Consider a product $p$. At a given point in time where the time index $t$ runs from 1 to $T$, we can draw a network, $N_{p,t}$, where the nodes are the countries and the edges represent the trade flows.
For any two countries $i$ and $j$, an edge is drawn from $j$ to $i$ if the former exports product $p$ at time point $t$ to the latter. Hence, the network is directed and  weighted. The weight of each edge is proportional to the monetary value obtained from the Comtrade database.

We investigate the following hypothesis of anti-dumping circumvention by \textit{re-routing} via a third country.
\begin{hyp} \label{hyp_rout}
Assuming that there is an anti-dumping duty on product $p$ on the trade flow going from country $j$ to country $i$ at time $t$, 
there is circumvention of the anti-dumping duty via re-routing if country $j$ re-routes product $p$ to country $i$ via  countries in the networks $N_{p, t}$ other than $i$ or $j$.
%Suppose Europe imposes an anti-dumping on product $p$ coming from country $A$ at time points $t$. Then we expect country $A$ to re-route product $p$ via other countries in the networks $N_{p, t}$. 
\end{hyp}

\begin{figure}
    \centering
    \includegraphics[scale=0.5]{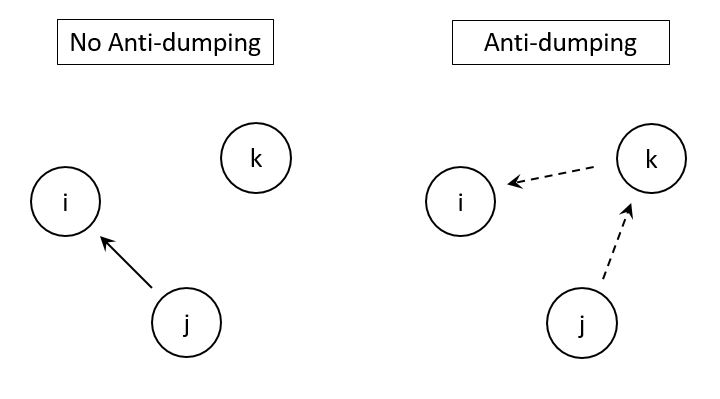}
    \caption{Expected  changes in trade flows between country $i$ and $j$ with no anti-dumping (left) and when, in presence of an anti-dumping, country $j$ circumvents the anti-dumping imposed by $i$ by re-routing via a third country $k$ (right).}
    \label{network_re-rout}
\end{figure}

To investigate the occurrence of circumvention of the anti-dumping duty via re-routing, we consider the network of trade flows of product $p$ at different time points $t$. 
Suppose an anti-dumping is imposed by country $i$ on country $j$ for product $p$. 
Under Hypothesis \ref{hyp_rout}, we expect country $j$ to re-route at least part of the trade flows of product $p$ via at least one third country, suppose country $k$.  
Figure \ref{network_re-rout} shows a stylized representation of anti-dumping circumvention, where the trade patterns between $i$ and $j$ have changed due to rerouting through country $k$.
We can quantify the extent to which re-routing takes place by computing all possible paths connecting country $j$ to $i$.  
When an anti-dumping is in place, we expect that a shift in the distribution of the weight of all possible shortest paths connecting $j$ to $i$ via any third country, due to re-routing. 
In particular, we expect an increase in the weight of paths of length greater than 1 connecting $j$ to $i$. 

%Instead of tracking the changes over time, we could also compare  changes in %two aggregated networks: one network where trade flows are aggregated over %time points when no anti-dumping on product $i$ in country A is into place, %one aggregated over time points when the anti-dumping is into place.

% Let $y^{p}_{A,t}$ be the median length of all possible paths connecting country $A$ to the EU at time $t$ for product $p$. We can consider the following equation for $y^{p}_{A,t}$: 
% \begin{equation}
%     y^{p}_{A,t} = antidump^{p}_{t} + W_{A,t} + \varepsilon^{p}_{A,t} 
% \end{equation}
% where $antidump^{p}_{t}$ is an indicator variable equal to 1 if at time $t$ there is an anti-dumping duty in force for product $p$ imported from country $A$, $W_{A,t}$ is a set of characteristics of country A observed at time t (e.g., including its geographical distance from the EU, the number of journal news that mention anti-dumping circumvention activities for country A, etc.), and $\varepsilon^{p}_{A,t}$ is an error term.

%If re-routing is observed, we can detect the re-routing countries, such as country B in Figure \ref{network_re-rout}, by computing ``betweenness" measures (see Table 5.8 in book by Bart Baessens). %In particular, we built five different measures based on the observed trade flow network that might indicate the potential circumvention of an anti-dumping duty.
Define $x_{i \rightarrow j, k}^{p,t}$ as the total of USD exchanged from country $i$ to country $j$ via a third country $k$ for product $p$ in period $t$: 
we obtain $x_{i \rightarrow j, k}^{p,t}$ by summing the values of the trade flows going from country $i$ to $k$ and from country $k$ to $j$, which we obtain from the Comtrade database. 
The measure of anti-dumping circumvention is obtained by summing $x_{i \rightarrow j, k}^{p,t}$ across all third countries other than $i$ and $j$:
\begin{equation}
    y_{j \rightarrow i}^{p, t} = \sum_{k \neq i,j} x_{i \rightarrow j, k}^{p,t}.
    \label{eq_cirucmevention}
\end{equation}
%For each trade flow (from country $i$ to country $j$, for product $p$, in period $t$), compute $r^{t}_{i \rightarrow j,l,k}$, defined as the total of USD exchanged from country $i$ to $j$ on paths of length $l$ for product $k$ in period $t$.
In the remainder, we refer to $ y_{j \rightarrow i}^{p, t}$ as the \textit{circumvention index}.
Alternatively, one could compute the $y$ measure considering other statistics than the monetary value of the trade flow, like the exchanged quantity (in Kg), or accounting only for the last section of the route (i.e., only incoming edges into the EU).
We compute the above measure on all routes between two trade partners that happen via a third country.
Moving to longer paths via more than one country would allow controlling for possible re-routing paths happening via more than one country: although this could potentially bring many additional insights to the analysis, it would also increase exponentially the computational complexity of the problem. 
Given the exploratory nature of this work, we consider the simplest case of circumvention happening only via one other country\footnote{The data set is available on the corresponding author's personal website {https://lucabarbaglia.github.io/}.}.

% \begin{enumerate}
% \item $n^\text{RR}_{i \rightarrow j,l,k,t}$  : number of paths of length $l$ going from country $i$ to $j$ for product $k$ in period $t$;
% \item $p^\text{RR}_{i \rightarrow j,l,k,t}$  : total of USD exchanged from country $i$ to $j$ on paths of length $l$ for product $k$ in period $t$;
% \item $q^\text{RR}_{i \rightarrow j,l,k,t}$  : total of kg exchanged from country $i$ to $j$ on paths of length $l$ for product $k$ in period $t$;
% \iten $xp_{i\rightarrow j,l,k,t}^{RR}$ : total of USD exchanged from country $i$ to $j$ on paths of length $l$ for product $k$ in period $t$ considering only the last section of the route (i.e., only incoming edges into EU);
% \item $xq_{i\rightarrow j,l,k,t}^{RR}$ : total of kg exchanged from country $i$ to $j$ on paths of length $l$ for product $k$ in period $t$ considering only the last section of the route (i.e., only incoming edges into EU);
% \end{enumerate}
% where the super-script $RR$ refers to the re-rerouting hypothesis.

\subsection{Econometric method}

In this section, we briefly discuss the econometric methods that we employ to detect a suspicious case of circumvention of an anti-dumping measure. 
For a given time series $y_{j \rightarrow i}^{p}$ of length $T$,  we construct a panel of  controls $\boldsymbol{y}_{j \rightarrow i}^{p'}$, being a matrix of dimension $T \times C_{p}$, where $C_p$ is the number of controls for product $p$. 
These controls correspond to the paths (of length 2) from country  $j$ towards country $i$ for all products $p'\neq p$ whose HS commodity codes have the first two digits different from the focal product $p$.
By considering products with HS code differing in the first two digits, we ensure that we have a large enough number of controls that belong to a different product category than product $p$.
Moreover, we select the control time series such that they are not associated with the announcement, imposition or withdrawal of any anti-dumping duty in the time period in analysis.
We exclude control time series if they have more than 10\% missing/non reported entries for the monetary value of the trade flow.
The number of controls $C_p$ varies across the 18 time series in analysis, ranging from a minimum of 121 controls to a maximum of 154 ones.\footnote{As a robustness check, we also looked at other control groups. For instance, we considered all the time series of the same product originating from all other countries than the one from the suspicious time series: the resulting number of controls was much smaller and results are not reported in the paper.}

%\paragraph{Fixed effects model for log-transformed response}
We pool the time series $y_{j \rightarrow i}^{p}$ and the respective controls $\boldsymbol{y}_{j \rightarrow i}^{p'}$ and estimate the fixed-effects regression model:
\begin{equation}
\text{log}(y_{j \rightarrow i}^{c,t}) = \mu^c + \beta d^{c,t} + \epsilon^{c,t},
\label{eq_panel}
\end{equation}
where 
%$y_{j \rightarrow i}^{c,t} \in [y_{j \rightarrow i}^{p,t},  \boldsymbol{y}_{j \rightarrow i}^{p',t}]$, and 
$c=1,\ldots, C_p+1$, and  $t = 1, \ldots, T$.
The index $c$ stands for either the focal product $p$ or one of the control products. 
We include a dummy  variable $d^{c,t}$ equal to one if time point $t$ is in-between (or including) the imposition and withdrawal of the anti-dumping duty for product $c$; zero otherwise.
The symbols $\mu^c$ and  $\epsilon^{c,t}$ stand for the fixed effect and error term, respectively.
The response is log-transformed, which is possible since it contains no zeros. 
Note also that the dummy $d^{c,t}$  takes  the value one only in the periods in-between the imposition and withdrawal of an anti-dumping duty of the time series in analysis $y_{j \rightarrow i}^{p}$. 
Indeed, the dummy $d^{c,t}$ is zero for all controls $\boldsymbol{y}_{j \rightarrow i}^{p'}$ by construction, since they are not subject to any anti-dumping duty.  
We test whether the estimated $\beta$ is significantly different from zero. %Table \ref{panel_log} presents the results. 
We also consider an alternative specification of Equation \eqref{eq_panel}, where we standardize the dependent variable by demeaning and rescaling.

% \subsection{Fixed effects model for response in standardized log-transformed response}
% For each of the $s=1, \ldots, S=18$ suspicions time series and their respective $C_s$ controls (where the number of controls thus depends on the suspicious time series), I estimate the fixed-effects panel data model:
% $$
% \tilde{y}_{it} = \mu_i + \beta d_{it} + \epsilon_{it}
% $$
% where $\tilde{y}_{it}$ is the standardized log-transformed response and $t=1,\ldots, 72$.
% I investigate whether $\beta$ is significantly different from zero. Table \ref{panel_logdiff} presents the results. 
% We consider two 

\section{Results\label{sec_results}}

\begin{figure}
    \centering
    \includegraphics[scale=0.75]{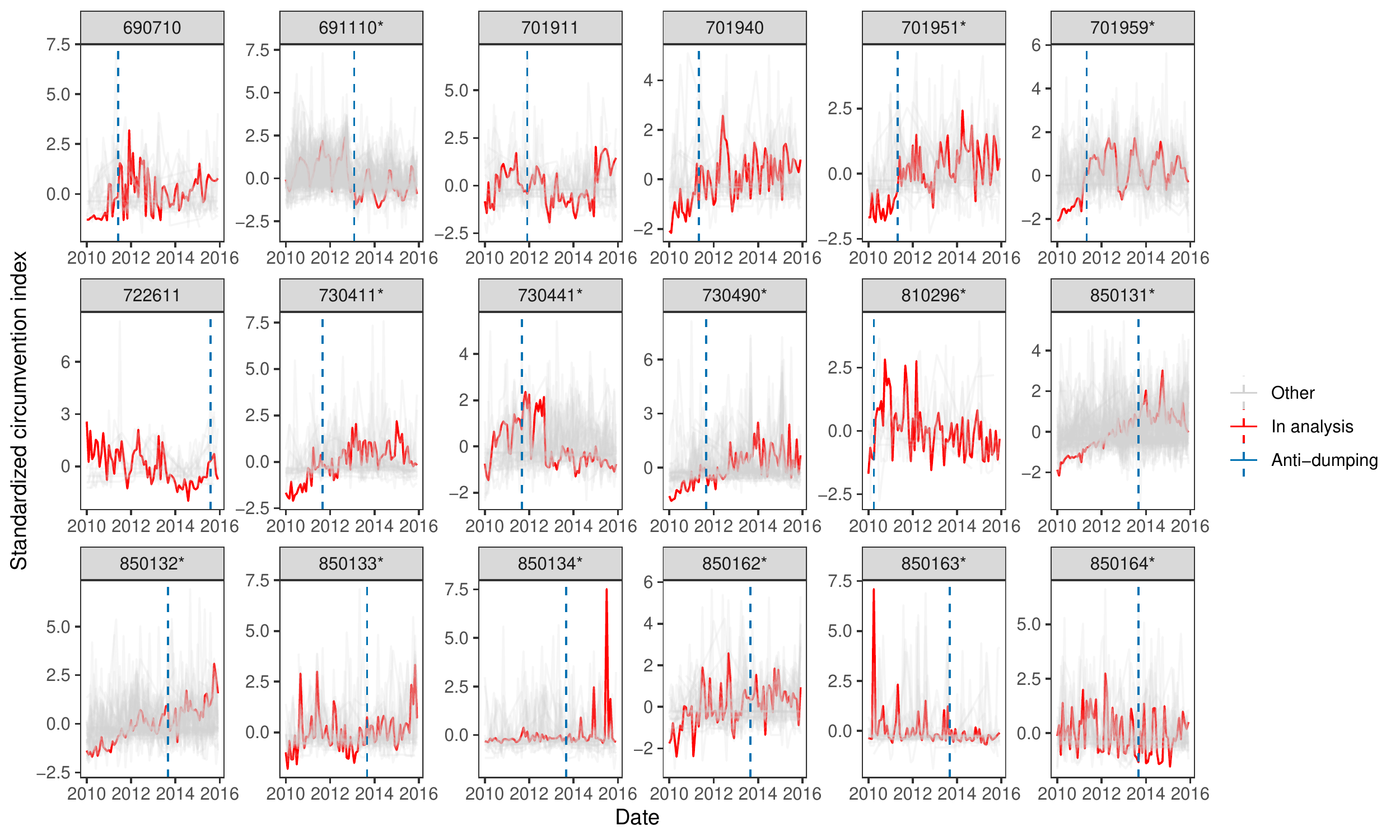}
    \caption{Standardized circumvention index for the 18 products in analysis: in red  the time series in analysis, in grey the circumvention indexes from other countries. 
    The dashed vertical line corresponds to the imposition of the anti-dumping duty on the series in analysis. 
    A * next to the commodity code indicates that the European authorities opened an investigation for anti-dumping circumvention.}
    \label{fig_circumention_index}
\end{figure}

Figure \ref{fig_circumention_index} reports the standardized \textit{circumvention index} defined in Equation \eqref{eq_cirucmevention} for each of the 18 time series in analysis. 
We report the circumvention index in analysis (in red), alongside  the circumvention indexes computed for all other countries for the same product (in grey). The  blue dashed vertical line indicates the date of imposition of the anti-dumping duty from the European authorities.
If there is circumvention of the anti-dumping duty via re-routing, we expect to observe an increase in the circumvention index of the time series under analysis. 
Such increase could have the form (i) of one or more unexpected jumps in the circumvention index under analysis, as seems to be the case for products like $850134$, or (ii) of a more stable increase in the circumvention index level, as seems to be the case for $701951$.

For measuring the performance of the proposed method for detecting anti-dumping circumvention we compute confusion matrices, see Table \ref{panel_log}. 
Recall that we conclude that there is suspicion of anti-dumping circumvention for the trade flow in analysis if the estimated coefficient $\hat{\beta}$ in Equation \eqref{eq_panel} is significantly different from zero.
% Table \ref{panel_log} reports the confusion matrix resulting from testing the significance of $\hat{\beta}$ in Equation \eqref{eq_panel}.
% If the estimated coefficient $\hat{\beta}$ is significantly different from zero, then  we conclude that there is suspicion of anti-dumping circumvention for the trade flow in analysis. 
The confusion matrix reports the number of cases where our methodology detects the anti-dumping circumvention (\textit{detected}) or not (\textit{not detected}), against the actual presence of an investigation about anti-dumping circumvention opened by the European authorities.
The off-diagonal entries of the confusion matrix report the incorrect classifications.
We report the result for different significance levels of the statistical test for detecting the null hypothesis of a change in the circumvention index. 
As expected, the lower the significance level, the harder for $\hat{\beta}$ to be significantly different from zero, thus decreasing the number of true positives (i.e., detected and under investigation) and false positives (i.e, detected and not under investigation).
% For both significance levels $0.1$ and $0.05$, we observe a worse performance in detecting true positives than the univariate case. 
% On the other hand, the performance at the $0.01$ significance level is better than the one reported in Table \ref{panel_log}. 
Considering a significance level of $\alpha=0.05$, we correctly detect 10 out of the the 14 trade flows as suspicious cases of  anti-dumping circumvention. 
Looking at the off-diagonal entries for $\alpha = 0.05$, we incorrectly detect two trade flows, while we miss  to detect four trade flows that were actually put under investigation by the European authorities.
%The off-diagonal entries of the confusion matrix report the incorrect classifications: using univariate dummy regressions, we incorrectly detect 2 time series, while we miss  to detect 3 trade flows that were actually put under investigation by the European authorities.

%% STANDARDIZE
Now we repeat the analysis using the standardized version of the circumvention index. Results are reported in Table \ref{panel_logdiff}. 
We observe an improvement at all significance levels.
%If we take the standardized circumvention index as the dependent variable in our panel model, we observe an improvement at all significance levels, as reported in Table \ref{panel_logdiff}.
In particular, at the standard  $0.05$ significance level, we correctly detect 86\% (12 out of 14) of the trade flows that have been subjected to an investigation about anti-dumping circumvention by the European authorities. 

\begin{table}[]
	\caption{
	Confusion matrices for detecting anti-dumping circumvention using  panel regression. We report the number of time series in analysis where the method detects a change or not (rows), as well as whether there is an anti-dumping investigation (columns). Results for different significant levels.
	%Panel data model with log-transformed series: Distribution of the 18 time series based on whether a significant dummy has been detected or not (rows) and the time series is under anti-dumping investigation or not (columns). Results for different significant levels  $\alpha=[0.1, 0.05, 0.01]$.
	} \label{panel_log}
	\centering
	\bigskip
	\begin{tabular}{rrrr}
		\hline
		Sign. level	&& Under Investigation & Not Under Investigation \\ 
		\hline
$\alpha=0.1$ &Detected &   11 &  2 \\ 
&Not Detected &   3 &  2 \\ 
		\hline
$\alpha=0.05$&Detected &  10 &   2 \\ 
&Not Detected &   4 &  2 \\ 

\hline
$\alpha=0.01$ &Detected &  9 &   2 \\ 
&Not Detected &   5 &   2 \\ 
\hline
	\end{tabular}
\bigskip
\bigskip
\end{table}

\begin{table}[]
	\caption{
	Confusion matrices for detecting anti-dumping circumvention using  panel regression with standardized response. We report the number of time series in analysis where the method detects a change or not (rows), as well as whether there is an anti-dumping investigation (columns). Results for different significant levels.
	%Panel data model with  series in standardized log-transformed series: Distribution of the 18 time series based on whether a significant dummy has been detected or not (rows) and the time series is under anti-dumping investigation or not (columns). Results for different significant levels  $\alpha=[0.1, 0.05, 0.01]$.
	} \label{panel_logdiff}
	\centering
	\bigskip
	\begin{tabular}{rrrr}
		\hline
		Sign. level	&& Under Investigation & Not Under Investigation \\ 
		\hline
$\alpha=0.1$&Detected &   12 &   2 \\ 
&Not Detected &  2 &  2 \\ 
		\hline
$\alpha=0.05$&Detected &   12 &   2 \\ 
&Not Detected &  2 &  2 \\ 
\hline
$\alpha=0.01$&Detected &   10 &   2 \\ 
&Not Detected &  4 &  2 \\ 
\hline
	\end{tabular}
\bigskip
\bigskip
\end{table}

\section{Conclusion\label{sec_conclusions}}

%% INTRO & OUR WORK
Anti-dumping duties are one of the common trade protection measures used by the EU to protect its member states' national economies.
Anti-dumping circumvention happens when a country attempts to avoid paying the duty imposed by the commercial partner, for instance, by  re-routing the trade flow via a third country.
In this work, we study possible events of anti-dumping circumvention via re-routing on a set of products imported by the EU.
Based on the observed network of trade flows among 205 partnering countries reporting bi-directional commercial exchanges, we build a \textit{circumvention index} to map possible events of re-routing via a third country present in our sample. 
We then adopt a panel method strategy on the proposed circumvention indexes to classify whether any of the trade flows in analysis might be flagged as a suspicious case of anti-dumping circumvention. 
We assess the validity of our empirical strategy against the actual investigations opened by the European authorities: our results indicate that the proposed empirical strategy is able to correctly identify 12 out of 14 cases under investigation for anti-dumping circumvention. 

%% FUTURE RESEARCH
Although exploratory in nature, this paper proposes a strategy to detect suspicious cases of anti-dumping circumvention relying on the observed network of trade flows. 
Such an approach might serve as an early-warning system for investigation authorities, who might use it to have a first screening of all trade flows directed to their countries, and then focus only on the suspicious cases to run a thorough investigation.
Future research should try to implement alternative empirical strategies to detect suspicious cases of anti-dumping circumvention, for instance adopting an approach in line with synthetic control methods (for a recent review, see \citealp{abadie2019using}).
Another interesting line for future work relates to studying circumvention of an anti-dumping duty via product mis-classification \cite{vermulst2015eu}, rather than re-routing via a third country.

\newpage
%\bibliographystyle{plainnat}
%\bibliography{bibliography.bib}
\bibliography{main.bbl}

\end{document}